\documentclass[review]{elsarticle}
\usepackage{graphicx,amsmath,amssymb,color,multirow,CJK,doi,ulem, array}

\begin{document}
\begin{frontmatter}

\title{Multiresolution community analysis of international trade networks}

\author{Wonguk Cho}
\address{Department of Physics, Sungkyunkwan University, Suwon 16419, Republic of Korea}
\address{Graduate School of Data Science, Seoul National University, Seoul 08826, Republic of Korea}
\author{Daekyung Lee}
\address{Department of Energy Technology, Korea Institute of Energy Technology, Naju 58322, Republic of Korea}
\author{Beom Jun Kim}
\ead{Corresponding author: beomjun@skku.edu}
\address{Department of Physics, Sungkyunkwan University, Suwon 16419, Republic of Korea}

\date{\today}

\begin{abstract}
The international trade network is a complex system where  multiple trade blocs with varying sizes coexist and overlap with each other. However, the resulting structures of community detection in trade networks are often inconsistent and fails to capture the complex landscape of international trade. To address these problems, we propose a multiresolution framework that aggregates all the configuration information from a range of resolutions. This allows us to consider trade communities of different sizes and illuminate the underlying hierarchical structure of trade networks and its constituting blocks. Furthermore, by measuring membership inconsistency (MeI) of each country and conducting multiple regression analysis with various economic and political indicators, we demonstrate that there exists a positive correlation between the external instability of countries and their structural inconsistency in terms of network topology.
\end{abstract}

\begin{keyword}
\texttt{Complex networks, Community detection, International trade, Hierarchical structure, Inconsistency}
\end{keyword}

\end{frontmatter}

\section{\label{sec:level1}Introduction}
The international trade is a complex system largely regulated by preferential trade agreements (PTAs), or trade blocs where a preferential access to certain products is given between the participating countries~\cite{united2020key}. In theory, any random pair of countries can sign a PTA with each other as long as it is beneficial for their economies. However, in reality PTAs do not spread in a uniform manner. The findings of Manger \textit{et al.}~\cite{manger2012hierarchy} suggest that structural arbitrage effects in trade networks provoke an endogenous spread of PTAs; some countries make a multitude of deals, others remaining relatively uninvolved.

In network theory we define a group of nodes that are more densely connected with nodes within the group as a community~\cite{newman2018networks, PhysRevE.69.026113}. A community detection method can be applied to find a community structure in trade networks and show which set of countries are more involved with each other. However, the conventional approach is fundamentally limited to the choice of a resolution parameter~\cite{arenas2008analysis, weir2017post, fortunato2007resolution}. That is, it can only identify a snapshot of trade communities from a certain range of sizes based on a given resolution. The dependence on the resolution parameter leads to a failure of reflecting the complex landscape of international trade, where multiple trade blocs with varying sizes coexist and overlap with each other.

On the other hand, applying community detection methods to trade networks is also susceptible to the inconsistency problem. That is, the resulting structures of community detection often show considerable variation, which is typically considered as evidence that the community detection methods are unreliable~\cite{kwak2009consistent, lancichinetti2012consensus, dahlin2013ensemble,gates2019element}. In order to address this issue, recent studies have suggested alternative approaches that focus on, instead of obtaining the single best division among high-score divisions, analyzing how similar those divisions are~\cite{PhysRevE.101.052306,kim2019relational, PhysRevE.103.052306}. Indeed, Riolo and Newman~\cite{PhysRevE.101.052306} demonstrate that seemingly inconsistent divisions can be assembled to identify a set of underlying subgroups of nodes (“building blocks”) that consistently share the same community membership across different divisions.

Furthermore, an attempt to quantify the extent of inconsistency in different divisions was made by Lee, \textit{et al}.~\cite{ PhysRevE.103.052306}, suggesting a method for measuring the inconsistency of community structures in each level of resolution. By applying their framework to different networks, they discover that the community structures of a resolution with high consistency not only remain relatively persistent when changing the resolution but also have a better representation of the true division.

In this paper, we address the importance of incorporating information from inconsistent divisions across different resolutions, as well as within each resolution. In real networks, particularly in international trade networks, each node belongs to multiple communities of different scales. The overlapping communities often lead to construct hierarchical structures where nodes that belong to the same community in a small scale also share the same community membership in a larger scale. Therefore, in order to fully understand the structural properties of nodes, we should examine the consistency of each pair of nodes throughout a range of scales.

Hence, we propose a multiresolution framework that incorporates information from inconsistent divisions across different resolutions. Our multiresolution framework consists of two major approaches: hierarchical decomposition analysis and membership inconsistency analysis. The former provides with a full description of the hierarchical structure of a network constructed upon its building blocks, while the latter elucidates the underlying structural inconsistency of a network in the individual node level.

Applying our multiresolution framework to trade networks, we obtain the “building blocks” of international trade, the set of countries that remain consistent irrespective of the resolution. Our method allows us to display the hierarchical structures built upon the building blocks, and the results accord with the regional blocks of the world economy. At the same time, we find the set of countries which are more inconsistent than others due to the structural characteristics they share. By multivariate regression analysis with various economic and political indicators, we demonstrate that there exists a positive correlation between the external instability of countries and their structural inconsistency in trade networks.

\section{\label{sec:level2}Methods}
\subsection{\label{sec:sub1}Multiresolution framework}
In this paper we use the generalized Louvain algorithm~\cite{jutla2011generalized}, a variant of the Louvain algorithm~\cite{blondel2008fast}, which detects the optimal division of a network by maximizing a generalized modularity function given by
\begin{equation}
Q = \frac{1}{2m}  \sum_{i,j} \left( A_{ij}-\gamma\frac{k_i k_j}{2m} \right), \delta(g_i,g_j),
\label{eq:Modularity general}
\end{equation}
where $\gamma$ is a resolution parameter, which can be adjusted to detect community structures in multiple scales. In our multiresolution framework, we define the resulting division from a realization of stochastic community detection algorithm as a configuration and the collection of configurations in each resolution $\gamma$ as an ensemble $C_\gamma$.

In theory, the possible choice of $\gamma$ ranges from zero to infinity. Therefore, it is necessary to decide a meaningful range of $\gamma$ or a resolution set $R$ for practical application. We first measure the average number of communities between configurations for each resolution, $ \langle n_\gamma \rangle$. The minimum resolution $\gamma_{\text{min}}$ can be given by the minimum level of $\gamma$ that satisfies $ \langle n_\gamma \rangle > 1$ because the range of resolutions where all nodes belong to a single community is irrelevant for the analysis of community structure.

On the other hand, at a high value of $\gamma$, the number of communities becomes nearly as many as the number of nodes. Clearly, configurations with the fragments of individual nodes are difficult to be considered as a meaningful representation of community structure. The appropriate limit of the smallest community scale would vary depending on the structural characteristic of a given network, but testing on various model networks, we find that setting $\gamma_{\text{max}}$ to be the maximum level of $\gamma$ that satisfies $ 3\langle n_\gamma \rangle \leq N$, where $N$ is the total number of nodes, serves the goal of our framework.

\subsection{\label{sec:sub2}Hierarchical decomposition analysis}
The hierarchical decomposition analysis aims to aggregate configurations from different resolutions and elucidate the underlying structure of a given network. To begin with, we find the set of nodes with consistent co-memberships by measuring the cooccurrence $\phi_{ij}^\gamma $ or the proportion of configurations in the ensemble for resolution $\gamma$ where nodes $i$, $j$ are assigned to the same community:

\begin{equation}
\phi_{ij}^\gamma =\frac{1}{\vert C_\gamma \vert} \sum_{\alpha \in C_\gamma} \delta_\alpha (g_i,g_j), 
\label{eq:phi}
\end{equation}
where $\vert C_\gamma \vert$ is the number of configurations in the ensemble for resolution $\gamma$ and is equivalent to the number of realizations for which the community detection algorithm runs at a given resolution, $\delta_\alpha$ is the Kronecker delta for configuration $\alpha$, and $g_i$ is the community membership index of node $i$. In Eq.~(\ref{eq:phi}), $\delta_\alpha (g_i,g_j)=1$ if nodes $i$, $j$ share the same community membership in configuration $\alpha$, and $\delta_\alpha (g_i, g_j)=0$ otherwise. Accordingly, the cooccurrence $\phi_{ij}^\gamma$ ranges from zero to unity. In order to extend this approach for different resolutions, we define the set of all ensembles from the range of resolutions as a multiresolution ensemble. Then, the multiresolution cooccurrence $\Phi_{ij}$, is given by
\begin{equation}
\Phi_{ij}=\frac{1}{\vert R \vert} \sum_{\gamma \in R}\phi_{ij}^\gamma,
\label{eq:Phi}
\end{equation}
where $\vert R \vert$ is the number of resolutions in the given resolution set $R$. The multiresolution occurrence $\Phi_{ij}$ measures how consistently nodes $i$, $j$ maintain their comembership throughout resolutions.

\begin{figure}[h!]
\includegraphics[width=0.98\textwidth]{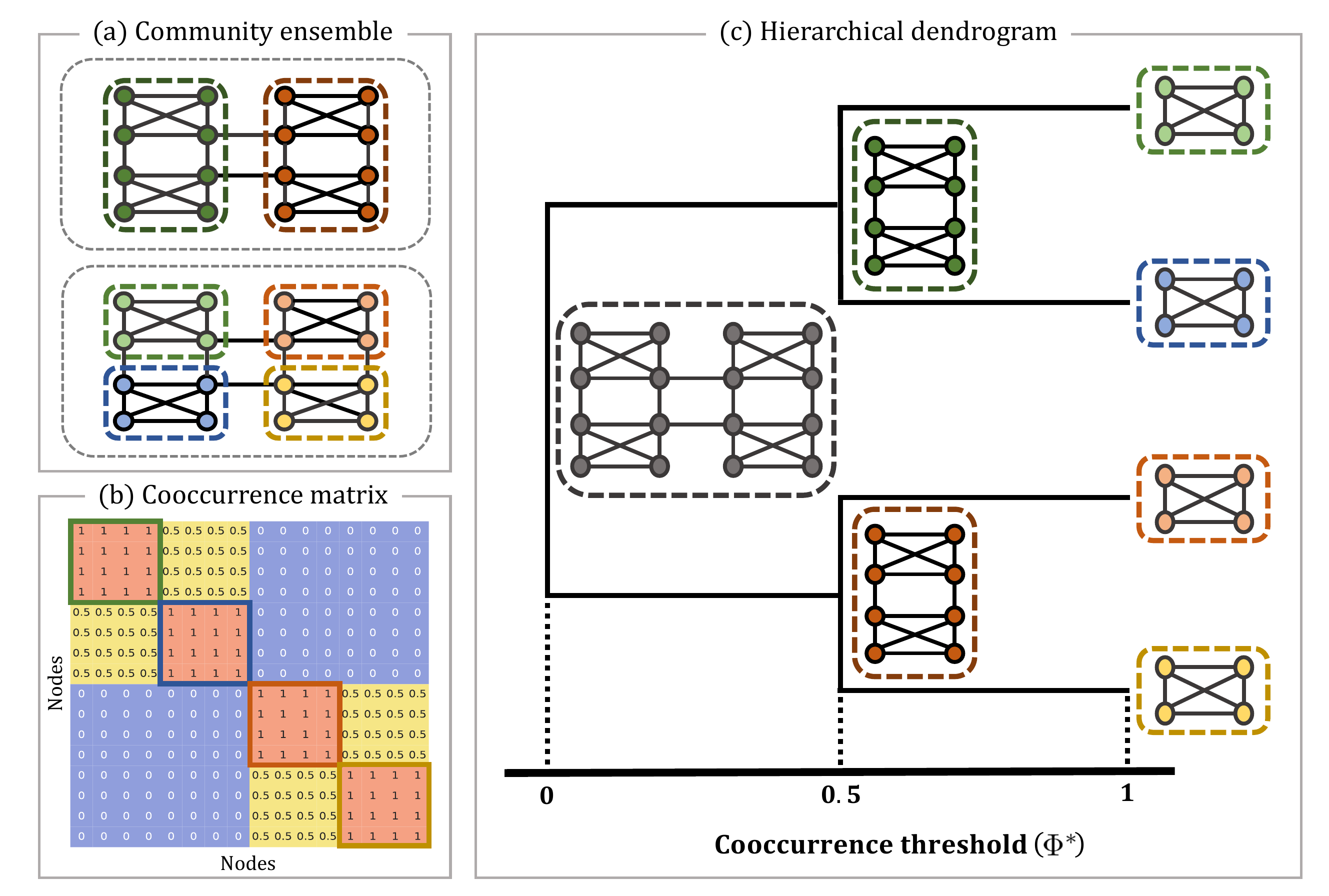}
\centering
\caption{An illustrative example of the hierarchical decomposition analysis. (a) The ensemble of the example network consists of two configurations, each accounts for $50\%$ of the ensemble. (b) The cooccurrence matrix is constructed from the ensemble. The blocks are colored red, yellow and blue for $\Phi_{ij}=1$, $0.5$ and $0$ respectively. (c) The dendrogram is built by the hierarchical decomposition. The horizontal axis refers to the cooccurrence threshold $\Phi^*$. For each merging point of branches, a configuration of the corresponding component is illustrated.}
\label{fig:decomposition}
\end{figure}

For instance, suppose a multiresolution ensemble that consists of two configurations shown in Fig.~\ref{fig:decomposition}(a); each of them accounts for $ 50\% $ of the ensemble. In the above configuration, two communities of eight nodes are detected, which are respectively subdivided into two communities of four nodes in the below configuration. The multiresolution cooccurrence matrix is shown in Fig.~\ref{fig:decomposition}(b), of which the elements are $\Phi_{ij}$. The red block diagonals represent the four communities in the below configuration, and their the members in each community consistently share the same community membership in the ensemble ($\Phi_{ij}=1$). The yellow-colored areas stand for the pairs of nodes that share the comembership only in the above configuration ($\Phi_{ij}=0.5$). The nodes that were never been assigned together are colored in blue ($\Phi_{ij}=0$).

The following step is to use the multiresolution cooccurrence matrix as an adjacency matrix to reconstruct the network into a complete weighted network where the weight of each edge corresponds to the cooccurrence $\Phi_{ij}$. Then, given a threshold $\Phi^* \in [0,1]$, we remove all the edges with the weight $\Phi_{ij} < \Phi^*$ from the reconstructed network and find a configuration of connected components. With $\Phi^*$ increasing from zero to unity, the fully-connected network becomes gradually decomposed into smaller components.

As shown in Fig.~\ref{fig:decomposition}(c), we can illustrate the entire composition of a network with a dendrogram. The horizontal axis refers to the cooccurrence threshold $\Phi^*$, and the configuration of each corresponding component is presented for every merging point of branches. In this case, the components or blocks of nodes described on the dendrogram accord with the representative communities detected in a certain level of resolution.

However, it is not necessarily the case in more complex networks. The components of the dendrogram, particularly those with the highest level of $\Phi^*$, may not correspond to communities themselves. Indeed, they are the "building blocks" of community structure which can be put together to assemble the detected communities. These blocks of nodes maintain their comemberships with high consistency throughout the change of resolutions and can be assembled to constitute all the communities from an ensemble. The dendrogram provides a description of the hierarchical structure the building blocks construct.

\subsection{\label{sec:sub3}Membership inconsistency analysis}
In the preceding section a method for decomposing a network into blocks of nodes based on their consistencies in a multiresolution ensemble is described. In this section we extend our multiresolution framework to analyze the structure of a network in the individual node level by measuring the consistency of communities for each node.

In order to quantify the node-level consistency, we employ a measure called membership inconsistency (MeI) proposed by Lee, \textit{et al.}~\cite{PhysRevE.103.052306}. First, let us denote a set of nodes which share community membership with node $i$ in each configuration $\alpha$ by $\psi_{i \alpha}$ such that
\begin{equation}
\psi_{i \alpha}= \{ j \mid g_{j \alpha} = g_{i \alpha} \}
\label{eq:psi}
\end{equation}
where $g_{i \alpha}$ refers to the community to which node $i$ is assigned in configuration $\alpha$. The consistency of node $i$ depends on how similar $\psi_{i \alpha}$ is between all the configurations in a multiresolution ensemble. The similarity between sets $\psi_{i \alpha}$ and $\psi_{i \beta}$ is given by
\begin{equation}
J_{i, \alpha \beta}= \frac {\vert \psi_{i \alpha} \cap \psi_{i \beta} \vert} {\vert \psi_{i \alpha} \cup \psi_{i \beta} \vert}
\label{eq:Jaccard}
\end{equation}
which is the Jaccard similarity coefficient. The maximum value of $J_{i, \alpha \beta}$ occurs when $\psi_{i \alpha}$ is equivalent to $\psi_{i \beta}$, where $J_{i, \alpha \beta}=1$. However, note that $J_{i, \alpha \beta}$ cannot be zero, for $\psi_{i \alpha}$ and $\psi_{i \beta}$ always share at least one node, node i itself. Using the Jaccard coefficient, the MeI of node $i$ in a multiresolution ensemble is defined as
\begin{equation}
\Psi_i = \frac{1}{\vert R \vert} \sum_{\gamma \in R} \frac{1}{\vert C_\gamma \vert} \sum_{\alpha,\beta \in C_\gamma}J_{i, \alpha \beta}
\label{eq:Psi}
\end{equation}
where $R$ is a resolution set and $C_\gamma$ is a configuration ensemble for resolution $\gamma$. The minimum value of $\Psi_i$ is unity, where the community of node $i$ is perfectly consistent for all the configurations in an ensemble.

\begin{figure}[h!]
\includegraphics[width=0.98\textwidth]{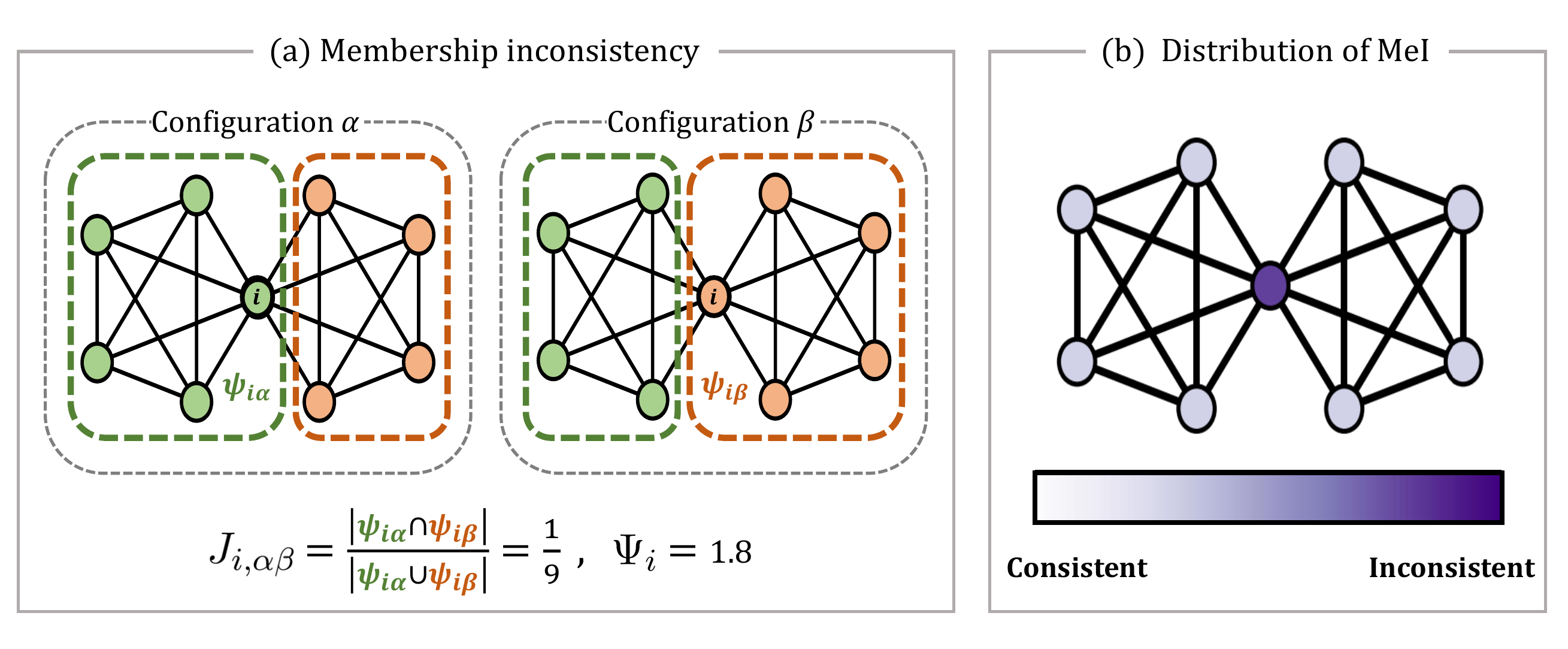}
\centering
\caption{An illustrative example of the membership inconsistency analysis. (a) The ensemble of the example network consists of two configurations $\alpha$ and $\beta$. From each configuration, the set of nodes that belong to the same community with node $i$ is given by $\psi_{i \alpha}$ and $\psi_{i \beta}$ respectively. (b) The distribution of $\Psi$ is illustrated on the network. The colors of the nodes indicate the relative strengths of MeI. A node with higher $\Psi$ is colored in darker purple.}
\label{fig:mei}
\end{figure}

An illustrative example of measuring the MeI value is given in Fig.~\ref{fig:mei}(a). In this network there are two cliques of five nodes, and the node in the middle (node $i$) constitutes the overlap of the cliques. Applying community detection algorithms to this network yields two different configurations where node $i$ alternates between two completely distinct communities ($\psi_{i \alpha} \cap \psi_{i \beta}=1$). The distribution of MeI is illustrated in Fig.~\ref{fig:mei}(b). The colors of the nodes indicate the relative strengths of MeI, and we can clearly notice that node $i$ in the middle shows the most inconsistent community membership.

\section{\label{sec:level3}Results}
\subsection{\label{sec:sub1}Hierarchical decomposition of trade networks}
In this study we use the historical trade data from the Trade Map database of International Trade Centre for the period 2010-2019~\cite{international2020trade}. We build an undirected weighted network of 45 countries based on the data. The directions of imports and exports are not considered because the meaning of communities in a directed network is rather unclear~\cite{malliaros2013clustering}, which is beyond the scope of the present study. Instead, we use annual bilateral trade volume as a weight between each pair of countries.

We apply the generalized Louvain algorithm~\cite{jutla2011generalized} for community detection to the international trade networks. The resolution parameter $\gamma_{\text{min}}$ is fixed at the minimum level of $\gamma$ that satisfies $ \langle n_\gamma \rangle > 1$. $\gamma_{\text{max}}$ is set to be the maximum level of $\gamma$ that satisfies $ 3\langle n_\gamma \rangle \leq N$. For each resolution, a community ensemble is formed by 1000 independent community detection results. By aggregating the configurations from the range of resolutions, we construct the multiresolution ensemble and conduct the hierarchical decomposition analysis by measuring the multiresolution cooccurrence $\Phi_{ij}$ in Eq.~(\ref{eq:Phi}).

\begin{figure}
\includegraphics[width=0.98\textwidth]{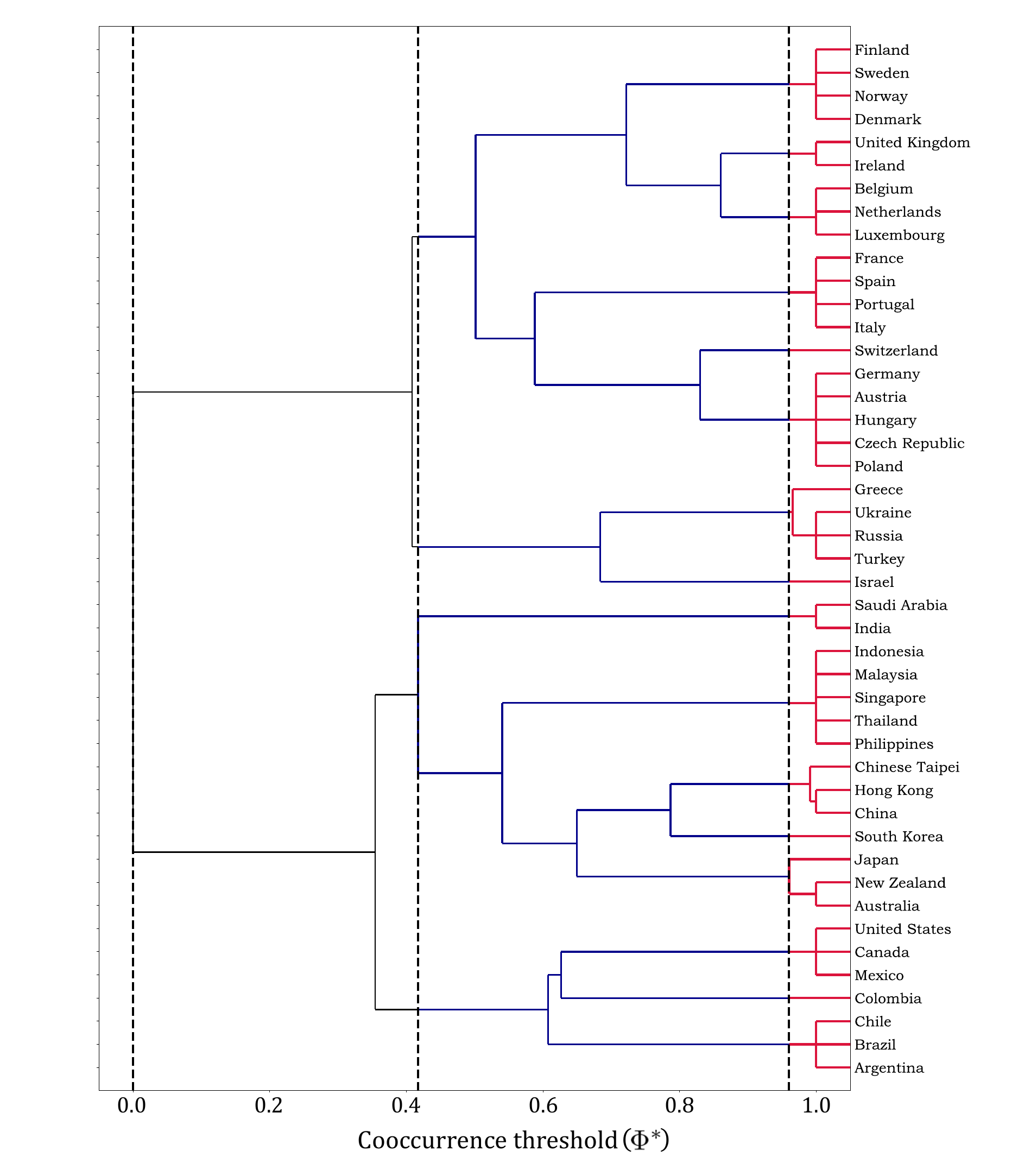}
\centering
\caption{The dendrogram of the international trade network in 2018 constructed by the hierarchical decomposition analysis. The horizontal axis refers to the cooccurrence threshold $\Phi^*$. The vertical dashed lines are drawn for $\Phi^*=0$, $0.41$ and $0.96$.}
\label{fig:dendrogram}
\end{figure}

\begin{figure}
\includegraphics[width=0.98\textwidth]{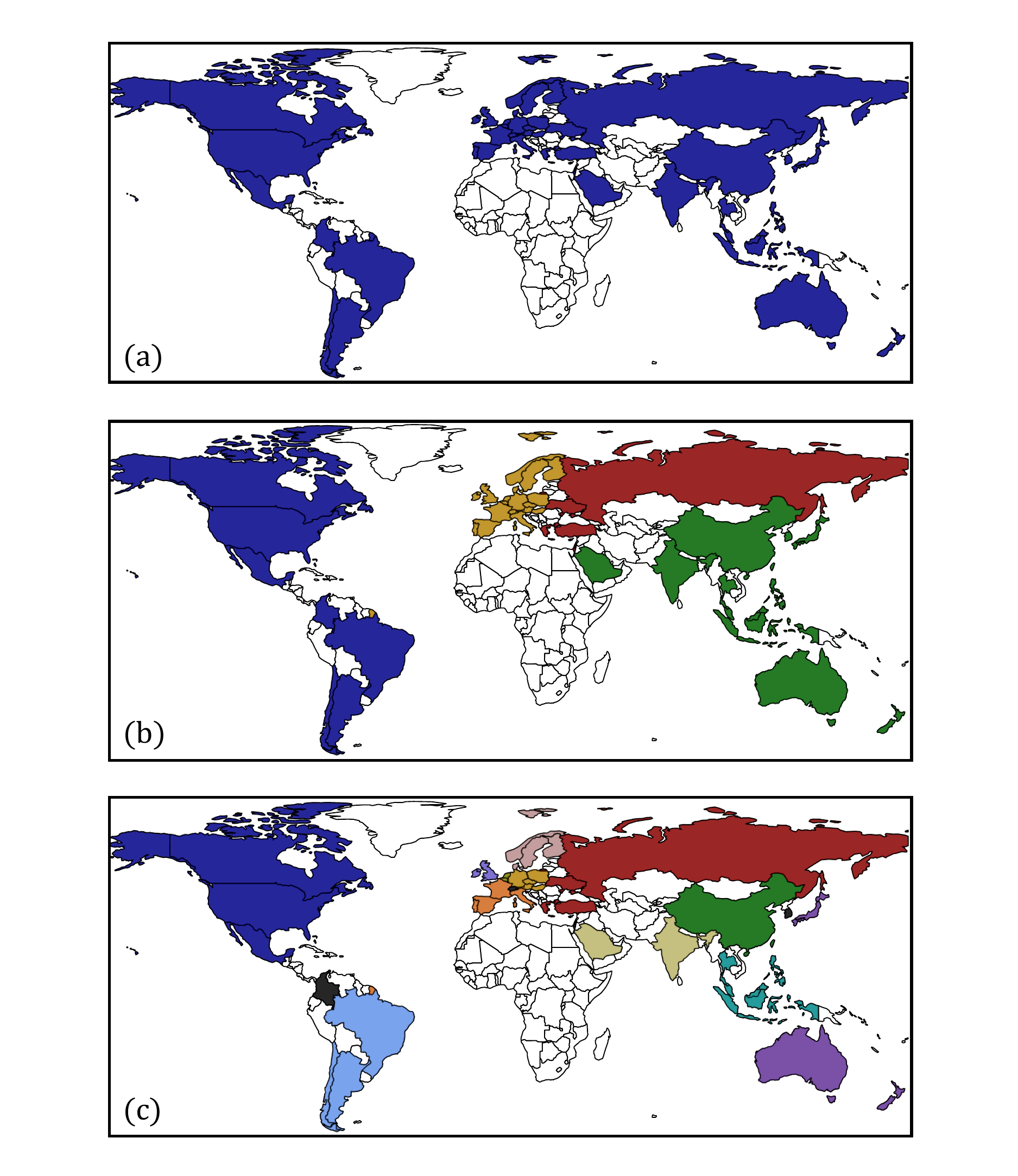}
\centering
\caption{The configurations of countries at the selected cooccurrence thresholds illustrated by the world map. (a) $\Phi^*=0$. (b) $\Phi^*=0.41$. (c) $\Phi^*=0.96$.}
\label{fig:map}
\end{figure}

The result of the hierarchical decomposition analysis is presented in Fig.~\ref{fig:dendrogram} and Fig.~\ref{fig:map}. Figure~\ref{fig:dendrogram} is a dendrogram that describes how the international trade network splits into smaller components with an increase of the cooccurrence threshold $\Phi^*$. Figure~\ref{fig:map}(a)-(c) illustrate the maps of divisions at $\Phi^*=0$, $0.41$, $0.96$ respectively. Beginning as a single large component at $\Phi^*=0$, the countries are divided into four components at $\Phi^*=0.41$ based on the continent: Western Europe, Eastern Europe, Asia and America. After multiple steps of splits, the countries are partitioned into 16 groups at $\Phi^*=0.96$. Each group consists of the countries that consistently sustained the same community membership over the range of resolutions. These groups are the building blocks of the international trade network. Note that these resolution-invariant building blocks accord with the regional blocs of the global economy. This implies that despite the deepening integration of world economies through globalization, the trading system still largely relies on the regional economic communities.

\subsection{\label{sec:sub2}Membership inconsistency of trade networks}
\begin{figure}
\includegraphics[width=0.98\textwidth]{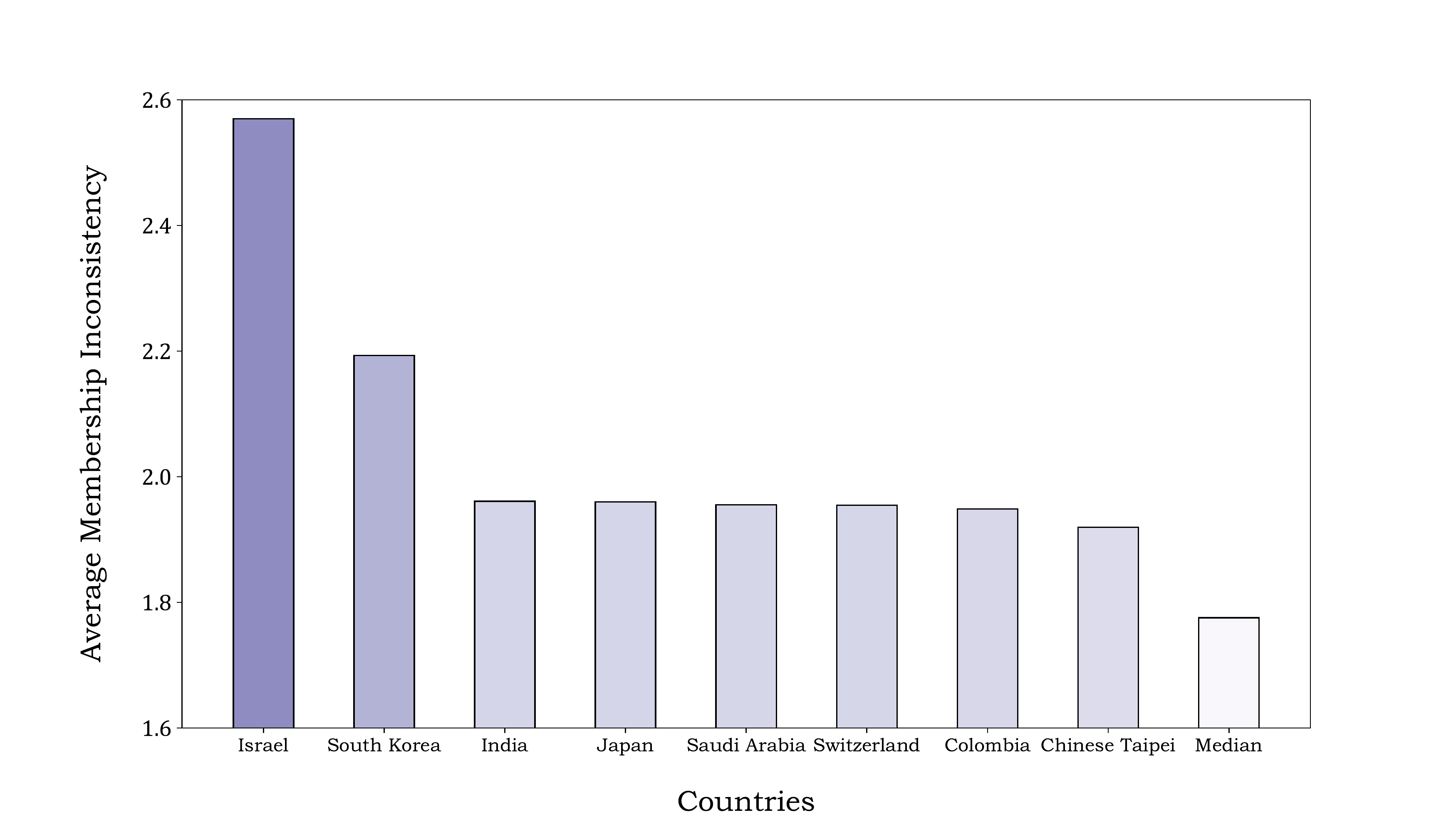}
\centering
\caption{The bar chart illustrates the average membership inconsistency (MeI) of the countries with the highest MeI and the median. The MeI is averaged for the period 2010-2019.}
\label{fig:bar}
\end{figure}

By using the multiresolution ensemble, we measure the membership inconsistency (MeI) of each country in Eq.~(\ref{eq:Psi}). The MeI value $\Psi$ in trade networks is a structural attribute of each country that represents the inconsistency between the compositions of trade communities that it belongs to. For example, suppose that a country is classified to be the same group with the United States in one configuration and with China in the other. If there exist few intersections between those two groups, the country will exhibit a high level of $\Psi$. The countries with the highest $\Psi$ are presented in Fig.~\ref{fig:bar}. In particular, Israel and South Korea have shown distinctly high $\Psi$ throughout a decade.

Although we can interpret the meaning of MeI in terms of network topology and find the set of countries with the higher structural inconsistency than others, it is unclear what this measure possibly implies in a real context. In previous literature, an attempt to find the link between network topological characteristics of a system and its physical properties was made by Kim \textit{et al.}~\cite{kim2015community}. With the application of community detection methods on the Chilean power grid, they find that the community consistency, a similar measure to MeI that quantifies how consistent community detection results are, determines the contribution of a node to the stability of electricity supply of a power grid.

In the present study, we propose a hypothesis that MeI influences the instability of a country in terms of external intervention. The rationale behind the hypothesis is that multiple membership of economic communities can put a country in a situation where it is confronted by conflicting interests between the communities~\cite{panke2018exploring}. In particular, considering a recent trend of international politics where trade is being weaponized as an explicit tool of strategic and political influence~\cite{harding2017weaponization}, overlapping memberships can result in instability of diplomatic relations. For instance, South Korea, which exhibits high $\Psi$,  has both important economic ties with China and a critical alliance with the United States. Hence, as the US-China rivalry intensifies, South Korea has to deal with political pressures from the two largest powers and struggles to determine its strategic direction for trade and foreign policy~\cite{sohn2019south}.

To test our hypothesis, we conduct a multiple regression analysis of MeI along with other political and economic factors. As a measure for instability from external intervention, we use the \textit{External Intervention Indicator} (EII) published by the Fund for Peace~\cite{haken2020failed}. The EII estimates the influence and impact of political and economic engagement of external actors in functioning of a country, and its score ranges from 0 (the least influenced) to 10 (the most influenced). In addition, three control variables are considered in estimating the impact of MeI (independent variable) on EII (dependent variable). These include the following:
\begin{itemize}
\item{\textit{GDP}}: Economic wealth of a country is intimately connected with its power and status in contemporary politics and functions as a powerful ordering principle in international governance~\cite{fioramonti2016post}. A standard measure of a country's economic wealth is Gross Domestic Product or GDP, and countries with high GDP are expected to be more resistant to external intervention. We control for this effect by including annual GDP (in trillions of current U.S. dollars) from the World Bank database~\cite{world2020world}.
\item{\textit{Political Stability}}: A country's political stability is an important factor that influences its foreign policy behavior~\cite{doi:10.1177/0022002714541842}, and unstable political situations often provide an opportunity for external intervention~\cite{pearson1974foreign}. We control for this effect by including the Political Stability Index (PSI) provided by the World Bank~\cite{ps2020}. The PSI is a composite measure that reflects the likelihood of a disorderly transfer of government power, social unrest, as well as ethnic, religious or regional conflicts, ranging from -2.5 (the most unstable) to 2.5 (the most stable).
\item{\textit{Trade Openness}}: High dependence on trade can make countries more vulnerable to external political pressure. In particular, the asymmetrical economic dependence between two partners compromises the foreign policy behavior of the more dependent~\cite{richardson1980trade}. We consider the effect of trade dependence by including trade openness (\% of GDP), which is defined as the ratio of exports plus imports over GDP~\cite{open2020world}.
\end{itemize}

\begin{table}[h!]
\label{regr}
\begin{center}
\caption{Regression results (Dep. variable: EII)}

\begin{tabular}{ | m{3.8cm}  m{2.6cm}   m{2.6cm}| } 
\hline
Variable & Coefficient    & Prob. ($P>|t|$)        \\  \hline
(Intercept) & \,\,1.899*** & $<0.001$\\
MeI & \,\,1.070***   &$<0.001$\\
GDP & -0.113***  & $<0.001$\\
PSI & -1.884***  &  $<0.001$\\ 
Trade Openness & \,\,0.001  &0.129\\\hline
\end{tabular}
\begin{tabular}{ | m{2.7cm}  m{1.587cm}  m{2.7cm}  m{1.587cm} |} 
{R$^2$}& 0.700  &  Prob.(F-statistic)  & $<0.001$\\
Adjusted R$^2$ & 0.697  & N & 430\\ \hline

\end{tabular}
\end{center}
\hspace{0.8cm} ***$p < 0.1\%$                                  
\end{table}

Table 1 presents the results of the regression model of the data for the period 2010-2019. The impact of MeI on EII is statistically significant ($ <0.1\% $), which corresponds to our hypothesis postulating that higher membership inconsistency in the community structure of trade networks leads to higher instability of a country in terms of external intervention. GDP and PSI also yield significant estimates as previously assumed. However, note that the effect of trade openness turns out to be not statistically significant. This is an unanticipated result as the dependence on trade is considered as an important factor that influences the vulnerability of countries to external shocks, particularly in contemporary politics where trade is being weaponized for strategic influence~\cite{harding2017weaponization,li2018economic}. One possible explanation is that simply measuring a share of trade in a country's economy without consideration of network topology fails to capture the complex interdependencies that affect economic and political behavior. This underlines the importance of understanding the community structure of trade networks and extracting the structural attributes of countries.

\section{\label{sec:level4}Summary and discussion}
In summary we propose a multiresolution framework that incorporates information from inconsistent divisions across different resolutions, as well as within each resolution. Applying our multiresolution framework to international trade networks, we present a hierarchical decomposition of trade communities and discover the building blocks of international trade, which accord with the regional blocks of the world economy. Besides, we measure the membership inconsistency of each country and find that it has a positive correlation with the external instability of countries.

The results of this study demonstrate that there exist important features of a network that can be identified only when all the resolutions are considered altogether. This new perspective can be further refined by testing in various community detection methods and algorithms. Furthermore, we have shown that the structural attributes in terms of network topology work as an effective measure to explain the existing measures in social science. From a practical view, the development of the new metrics for international trade opens up new possibilities to accurately evaluate the characteristics and behaviors of countries in both economic and political perspectives.

\section*{Acknowledgement}
This research was supported by the National Research Foundation of Korea (NRF) grant funded by the Korean government (MSIT), Grant No. 2019R1A2C
2089463.

\bibliographystyle{elsarticle-num}

\bibliography{main}
\end{document}